\documentstyle [prl,multicol,aps,psfig]{revtex}
\begin{document}
\title{Delocalization in the Anderson model due to a local measurement}
\author{S.A. Gurvitz}
\address{Department of Particle Physics,
Weizmann Institute of Science, Rehovot~76100, Israel}
\vspace{18pt}
\maketitle
\begin{abstract}
We study a one-dimensional Anderson model in which one site 
interacts with a detector monitoring the occupation of that 
site. We demonstrate that such an interaction, no matter how weak,  
leads to total delocalization of the Anderson model, and we 
discuss the experimental consequences.
\end{abstract}
\hspace{1.5 cm}  PACS: 03.65.Bz, 73.20.Fz, 73.20.Jc 
\vspace{18pt}
\begin{multicols}{1}
Consider an electron in a one-dimensional array of $N$ coupled wells.
The system is described by the Anderson tunneling Hamiltonian
\begin{equation}
H_A=\sum_{j=1}^NE_jc^\dagger_jc_j+\sum_{j=1}^{N-1}
(\Omega_j c^\dagger_{j+1}c_j+H.c.)\ ,
\label{a1}
\end{equation}
where the operator $c^\dagger_j\ (c_j)$ corresponds to the creation 
(annihilation) of an electron in the well $j$. 
We assume for simplicity that each of the wells contains one 
bound state $E_j$ and is coupled only to its nearest neighbors
with couplings $\Omega_j$ and $\Omega_{j-1}$. 
(We choose $\Omega_j$ real without loss of generality.)

The electron-wave function in this system can be written as
$|\Psi (t)\rangle =\sum_j b_j(t)c^\dagger_j |0\rangle$, where $b_j(t)$ 
is the probability amplitude of finding the electron in the well $j$
at time $t$. These amplitudes are obtained from 
the time-dependent Schr\"odinger equation $i\partial_t|\Psi (t)\rangle 
=H_A|\Psi (t)\rangle$.  
It is well known that for randomly distributed levels $E_j$ (or 
random couplings $\Omega_j$) all electronic states in this structure 
are localized\cite{ander}. Hence, if the electron initially occupies 
the first well, $b_j(0)=\delta_{j1}$, the probability of finding it  
in the last well, $P_N(t)=|b_N(t)|^2$ drops exponentially with $N$:
$\langle P_N(t\to\infty)\rangle_{ensemble}\to\exp(-\alpha N)$. 

Anderson localization is usually associated with destructive 
quantum-mechanical interference between different probability 
amplitudes $b_j(t)$. This interference, however, can be affected 
by measuring the electron's position in the system 
due to interaction of the electron with 
a macroscopic detector.  For instance, the continuous monitoring of one 
of the wells of a double-well system ($N=2$ in Eq.~(\ref{a1})) 
destroys the off-diagonal elements (coherences) 
of the electron density matrix. As a result, the latter become 
the statistical mixture: 
$\sigma_{jj'}(t)\to (1/2)\delta_{jj'}$ for $t\to\infty$\cite{gur1}. 

In the case of the $N$-well structure, however, the monitoring of one 
of the wells cannot determine the electron's position in the entire 
system. One might suppose therefore that such a local measurement cannot 
totally destroy the interference inside the entire system and hence,     
the electron localization. We demonstrate in this letter the contrary:
any interaction, no matter how weak, of the electron with 
a macroscopic detector placed on only {\em one} 
well leads to total delocalization of the electron state:
$P_N(t\to\infty)= 1/N$, even when $N\to\infty$. 

As a physical realization we consider a mesoscopic system of coupled 
quantum dots (Fig.~1), where a point contact is placed near the first dot. 
The point contact
\begin{figure} 
\psfig{figure=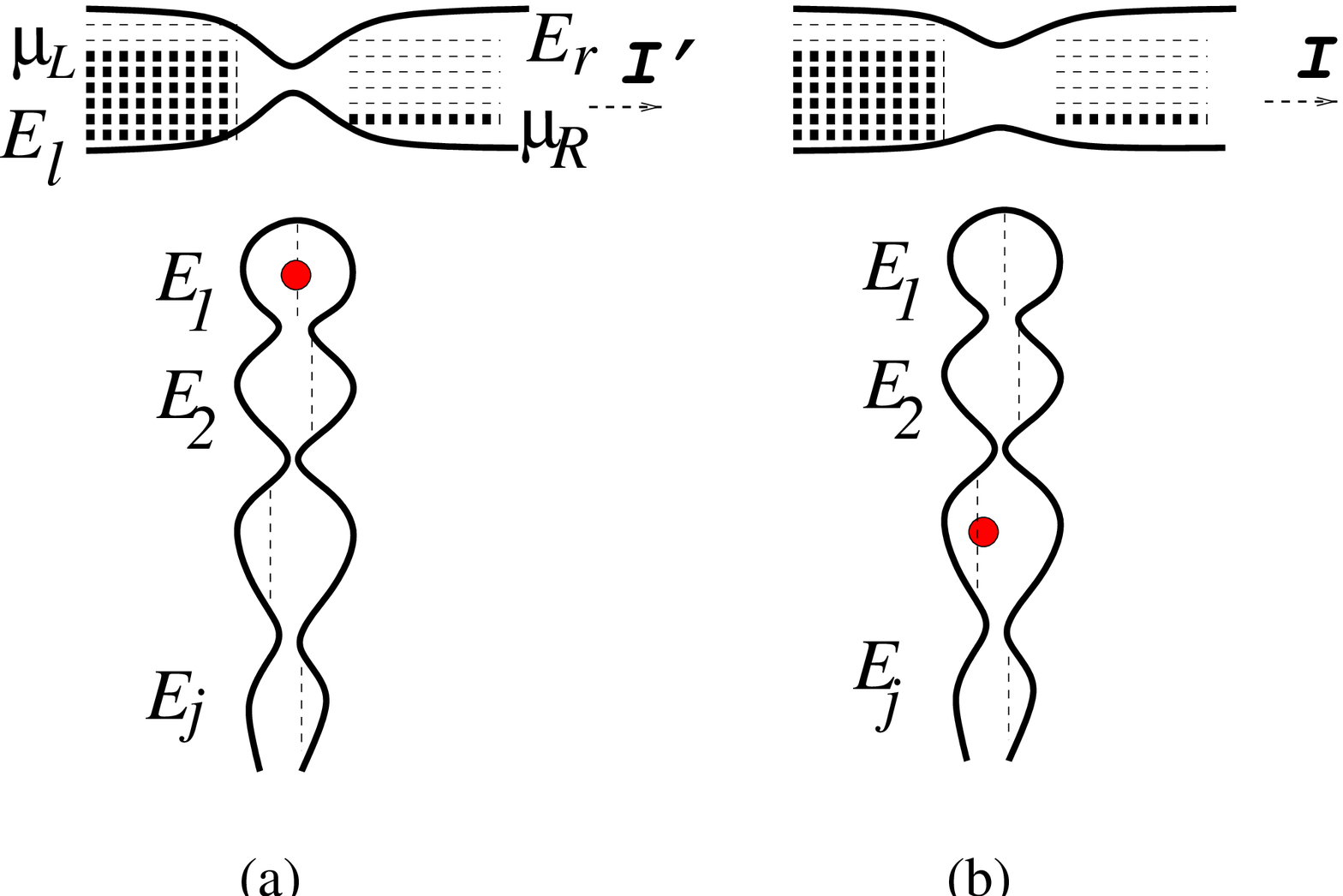,height=5cm,width=8.5cm,angle=0}
\noindent
{\bf Fig.~1:}
The point-contact detector near the array of coupled dots
with randomly distributed energy levels,  
described by the Anderson Hamiltonian (\ref{a1}). The detector 
monitors only the occupation of the first dot.
\end{figure}
\noindent
is coupled to two reservoirs, emitter and collector, at different chemical 
potentials, $\mu_L$ and  
$\mu_R$. A current $I=eT(\mu_L-\mu_R)/(2\pi)$ flows through 
the point contact\cite{land}, 
where $T$ is its transmission coefficient. If the electron occupies 
the first dot, the transmission coefficient of the point contact 
decreases, $T'<T$, due to the electrostatic repulsion 
generated by the electron. As a result, 
the current $I'<I$ (Fig.~1a). The current 
returns to its previous value $I$ whenever the electron occupies 
any other dot, since then it is far away from the contact (Fig.~1b).  

The entire system can be described by the tunneling 
Hamiltonian $H=H_A+H_{PC}+H_{int}$, where $H_A$ is given by  
Eq.~(\ref{a1}) and 
\begin{eqnarray}
H_{PC}&=&\sum_lE_la_l^\dagger a_l+\sum_rE_ra_r^\dagger a_r
+\sum_{l,r}(\Omega_{lr}a_r^\dagger a_l+H.c.)
\nonumber\\
H_{int}&=&\sum_{l,r}\delta\Omega_{lr}c^\dagger_1c_1
(a_r^\dagger a_l+H.c.),
\label{a3}
\end{eqnarray}
where $a_l^\dagger (a_l)$ and $a_r^\dagger (a_r)$ are the creation 
(annihilation) operators in the left and the right reservoirs, 
and $\Omega_{lr}$ is the hopping amplitude between the 
states $l$ and $r$ of the reservoirs. 

Consider an initial state where all the levels in the emitter 
and the collector are filled 
up to the Fermi energies $\mu_L$ and $\mu_R$, respectively,    
and the electron occupies 
the first well. The many-body wave function describing the entire 
system can be written in the occupation number representation as 
\begin{eqnarray}
&&|\Psi (t)\rangle =\sum_{j=1}^N\left [b_j(t)c_j^\dagger
+\sum_{l,r}b_{jlr}(t)c_j^\dagger a_r^\dagger a_l\right.\nonumber\\
&&\left.~~~~~~~~~~~+\sum_{l<l',r<r'}
b_{jll'rr'}(t)c_j^\dagger a_r^\dagger a_{r'}^\dagger a_l a_{l'}+\cdots
\right ]|0\rangle\ ,
\label{a4}
\end{eqnarray}
where $b(t)$ are the probability amplitudes of finding the system 
in the states defined by the corresponding creation and 
annihilation operators. Using these amplitudes one defines 
the reduced density matrices $\sigma_{jj'}^{(m)} (t)$ that 
describe the electron and the detector, 
\begin{eqnarray}
&&\sigma^{(0)}_{jj'}(t)=b_j(t)b^*_{j'}(t),~~~~
\sigma^{(1)}_{jj'}(t)=\sum_{l,r}b_{jlr}(t)b^*_{j'lr}(t),~~~~\nonumber\\
&&
\sigma^{(2)}_{jj'}(t)=\sum_{ll',rr'}b_{jll'rr'}(t)b^*_{j'll'rr'}(t),\; 
~\cdots\
\label{a5}
\end{eqnarray}
Here $j,j'=\{ 1,2,\ldots ,N\}$ denote the occupation states of 
the $N$-dot system. 
The index $m$ denotes the number of electrons that have reached 
the right-hand reservoir by time $t$. The total probability 
for the electron to occupy the dot $j$ is
$\sigma_{jj}(t)=\sum_m\sigma^{(m)}_{jj}(t)$. The off-diagonal 
density-matrix element $\sigma_{jj'}(t)=\sum_m\sigma^{(m)}_{jj'}(t)$
describes interference between the states $E_j$ and $E_{j'}$.

In order to find the amplitudes $b(t)$, we 
substitute Eq.~(\ref{a4}) into the time-dependent 
Schr\"odinger equation $i\partial_t|\Psi (t)\rangle 
=H|\Psi (t)\rangle$,  and use the
Laplace transform $\tilde b(E)=\int_0^\infty b(t)\exp (iEt)dt$.
Then we find an infinite set of algebraic equations for 
the amplitudes $\tilde b(E)$, given by 
\begin{mathletters}
\label{a6}
\begin{eqnarray}
&&(E-E_1) \tilde{b}_{1} - \Omega_1\tilde b_2 
-\sum_{l,r} \Omega'_{lr}\tilde{b}_{1lr}=i
\label{a6a}\\
&&(E-E_2) \tilde{b}_{2}  - \Omega_1\tilde b_1 -\Omega_2\tilde b_3 
- \sum_{l,r} \Omega_{lr}\tilde{b}_{2lr}=0
\label{a6b}\\
&&(E + E_{l}-E_1 - E_r) \tilde{b}_{1lr} - \Omega'_{lr}\tilde{b}_1
-\Omega_1\tilde b_{2lr}\nonumber\\ 
&&~~~~~~~~~~~~~~~~~~~~~~~~~~~~~~~~~
-\sum_{l',r'}\Omega'_{l'r'}\tilde{b}_{1ll'rr'}=0
\label{a6c}\\
&&(E + E_{l}-E_2 - E_r) \tilde{b}_{2lr} - \Omega_{lr}\tilde{b}_2
- \Omega_1\tilde{b}_{1lr} \nonumber\\
&&~~~~~~~~~~~~~~~~~~~~~~~ 
-\Omega_2\tilde b_{3lr}-\sum_{l',r'}\Omega_{l'r'}\tilde{b}_{2ll'rr'}=0
\label{a6d}\\
&&~~~~~~~~~~~~~~~~~~~~~~~~~~~~~~~~~~~\cdots\, , 
\nonumber
\end{eqnarray}
\end{mathletters}
where $\Omega'_{lr}=\Omega_{lr}+\delta\Omega_{lr}$.
 
Eqs.~(\ref{a6}) can be converted to Bloch-type equations for 
the density matrix $\sigma_{jj'}(t)$ without their explicit 
solution. This technique has been derived in
\cite{gur1,gur2}. We explain below only the main points of 
this procedure and the conditions for its validity.
 
Consider, for example, Eq.~(\ref{a6a}). 
In order to perform the summation in the term 
$\sum_{l,r} \Omega'_{lr}\tilde{b}_{1lr}$, we solve for $\tilde b_{1lr}$
in Eq.~(\ref{a6c}). Then substituting the result 
into the sum, we can rewrite Eq.~(\ref{a6a}) as 
\begin{eqnarray}
&&\left (E-E_1
-\int{{\Omega'}^2_{lr}\rho_L(E_l)\rho_R(E_r)dE_ldE_r
\over E+E_l-E_1-E_r}\right ) 
\tilde{b}_{1}\nonumber\\
&&~~~~~~~~~~~~~~~~~~~~~~~~~~~~~~~~~~~~~~~~- \Omega_1\tilde b_2 
+{\cal F}=i\ ,
\label{a7}
\end{eqnarray}
where we have replaced the sum in Eq.~(\ref{a6a}) by an integral 
$\sum_{l,r}\;\rightarrow\;\int 
\rho_{L}(E_{l})\rho_{R}(E_{r})\,dE_{l}dE_r\:$,
with $\rho_{L,R}$ the density of states in the emitter and collector. 
We split this integral into its principle value and singular part. 
The singular part yields $iD'/2$, where  
$D'=2\pi{\Omega'}^2\rho_L\rho_R (\mu_L-\mu_R)$, and 
the principal part is zero, providing $\Omega'_{lr}$ and 
$\rho_{L,R}$ are weakly dependent on the energies $E_{l,r}$. 
Note that $(2\pi)^2\Omega^2\rho_L\rho_R=T$\cite{bardeen},
where $T$ is the tunneling transmission coefficient of the point contact. 
Thus, $eD'=I'$ is the current 
flowing through the point contact\cite{land} whenever 
the electron occupies the first dot.

The quantity ${\cal F}$ in Eq.~(\ref{a7}) denotes 
the terms in which the amplitudes $\tilde b$ 
cannot be factored out of the integrals. 
These terms vanish in the large-bias limit,
$(\mu_L-\mu_R)\gg\Omega^2\rho$. 
Indeed, all the singularities 
of the amplitude $\tilde{b} (E,E_l,E_{l'},E_r,E_{r'})$
in the $E_l, E_{l'}$ variables lie below the real axis. 
This can be seen directly from Eqs.~(\ref{a6})
by noting that $E$ lies above the real axis in the Laplace 
transform. Assuming that the transition amplitudes 
$\Omega$ as well as the densities of states $\rho_{L,R}$ are
independent of $E_{l,r}$, one can close the integration contour 
in the upper $E_{l,r}$-plane. Since the integrand decreases 
faster than $1/E_{l,r}$, the resulting integrals are zero. 

Applying analogous considerations to the other equations of the
system (\ref{a6}) we convert Eqs.~(\ref{a6}) directly into 
rate equations via the inverse Laplace transform. 
The details can be found in
\cite{gur1,gur2,eg}. Here we present only the final equations 
for the electron density matrix $\sigma_{jj'}(t)$:
\begin{mathletters}
\label{a10}
\begin{eqnarray}
&&\dot\sigma_{jj}=i\Omega_{j-1} (\sigma_{j,j-1}-\sigma_{j-1,j})
\nonumber\\
&&~~~~~~~~~~~~~~~~~~~~~~~~~~~~~~~
+i\Omega_j (\sigma_{j,j+1}-\sigma_{j+1,j}),
\label{a10a}\\
&&\dot\sigma_{jj'}=i\epsilon_{j'j}\sigma_{jj'}+
i\Omega_{j'-1}\sigma_{j,j'-1}+i\Omega_{j'}\sigma_{j,j'+1}\nonumber\\
&&~-i\Omega_{j-1}\sigma_{j-1,j'}-i\Omega_j\sigma_{j+1,j'}
-\frac{\Gamma}{2}\sigma_{jj'}(\delta_{1j}+\delta_{1j'})\, ,
\label{a10b}
\end{eqnarray}
\end{mathletters}
where $\epsilon_{j'j}=E_{j'}-E_j$ and  
$\Gamma=(\sqrt{I/e}-\sqrt{I'/e})^2$ is the decoherence rate,
generated by interaction with the detector.
Note that these equations have been obtained 
from the many-body Schr\"odinger equation for the entire system. 
No stochastic assumptions have been made in their derivation.

Eqs.~(\ref{a10}) are analogous to the well-known optical Bloch equations
used to describe a multilevel 
atom interacting with the quantized electromagnetic field\cite{bloch}.
To our knowledge, this is their first appearance in connection with 
the Anderson model. The equations can be rewritten in 
Lindblad\cite{lind} form as 
\begin{equation}
\dot\sigma=-i[H_A,\sigma ]-{\Gamma\over 2}(Q\sigma+\sigma Q
-2\tilde Q\sigma \tilde Q^\dagger)\ ,
\label{lind}
\end{equation}
where $H_A$ is given by Eq.~(\ref{a1}) and 
$Q_{jj'}=\tilde Q_{jj'}=\delta_{1j}\delta_{1j'}$.
If $\Gamma=0$, Eq.~(\ref{lind})
is equivalent to the Schr\"odinger equation $i\partial_t|\Psi (t)\rangle 
=H_A|\Psi (t)\rangle$. In this case the electron 
density matrix $\sigma (t)$ displays Anderson localization, 
i.e., $\sigma_{NN}(t\to\infty )
\sim\exp(-\alpha N)$. If $\Gamma\not =0$, however, the 
asymptotic behavior of the reduced density-matrix, 
$\sigma_{jj'}(t\to\infty)$, changes dramatically: 
all eigenfrequencies (except for the zero mode)  
obtain an imaginary part due to the second 
(damping) term in Eq.~(\ref{lind}), so that only the stationary terms 
survive in the limit $t\to\infty$. This damping is illustrated in Fig.~2 
which displays the numerical
\begin{figure} 
\psfig{figure=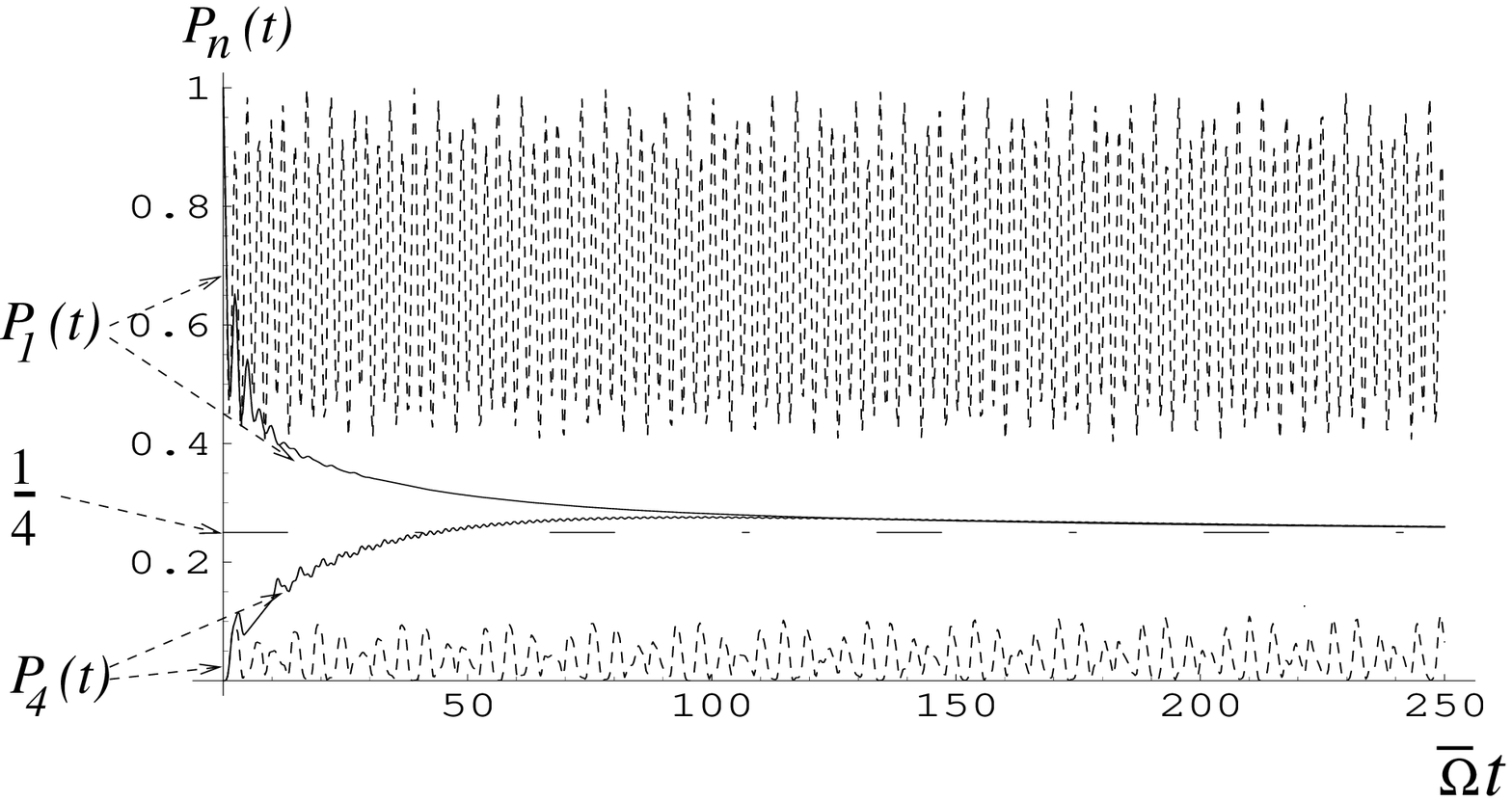,height=7cm,width=8.5cm,angle=0}
\noindent
{\bf Fig.~2:}
$P_1(t)$ and $P_4(t)$ represent the occupation of the first and 
the last dot as a function of time. 
The dashed lines correspond to $\Gamma=0$ (no interaction with 
the environment) and the solid lines correspond to $\Gamma/\bar\Omega =1$.
\end{figure}
\noindent
solution of Eqs.~(\ref{a10}) for $N=4$, $\Omega_j=\bar\Omega$=const, and 
$E_j/\bar\Omega =\{0,\ 2,\ 4, \ 1\}$. The occupation 
of the first dot, $P_1(t)=\sigma_{11}(t)$, and the last dot, 
$P_4(t)=\sigma_{44}(t)$, 
is shown in Fig.~2 for $\Gamma=0$ by the dashed lines, and for 
$\Gamma/\bar\Omega=1$ by the solid lines. One can clearly see  
that all oscillations decay for $\Gamma\not =0$, so that 
the density matrix reaches a stationary limit. Then we see the 
opposite of localization, as the
probability of finding the electron in the last dot, 
$P_4(t)$, becomes the same as the probability of finding it in 
the first dot, $P_1(t)$. 

The delocalization phenomenon, illustrated by Fig.~2, can be proven
analytically for any $N$. 
Indeed, let us consider Eqs.~(\ref{a10}) in the asymptotic limit 
$t\to\infty$, where the electron density matrix reaches its stationary 
limit: $\sigma_{jj'}(t\to\infty ) =u_{jj'}+i\ v_{jj'}$. 
Since for the stationary solution $\partial_t\sigma_{jj'}\to 0$, 
Eqs.~(\ref{a10}) become 
\begin{mathletters}
\label{a11}
\begin{eqnarray}
0& = & \epsilon_{j'j}v_{jj'}+
\Omega_{j'-1}v_{j,j'-1}+\Omega_{j'}v_{j,j'+1}
-\Omega_{j-1}v_{j-1,j'}\nonumber\\
&&~~~~~-\Omega_jv_{j+1,j'}
+\frac{\Gamma}{2}u_{jj'}(\delta_{1j}+\delta_{1j'})(1-\delta_{jj'})\ ,
\label{a11a}\\
0& = & \epsilon_{j'j}u_{jj'}+
\Omega_{j'-1}u_{j,j'-1}+\Omega_{j'}u_{j,j'+1}
-\Omega_{j-1}u_{j-1,j'}\nonumber\\
&&~~~~~-\Omega_ju_{j+1,j'}
-\frac{\Gamma}{2}v_{jj'}(\delta_{1j}+\delta_{1j'})(1-\delta_{jj'})\ .
\label{a11b}
\end{eqnarray}
\end{mathletters}
Eqs.~(\ref{a11}) have the unique solution 
$v_{jj'}=0$ and $u_{jj'}=(1/N)\delta_{jj'}$. 
This can be obtained by solving these equations sequentially, 
starting with $j,j'=N$, and then continuing for $j,j'=N-1,N-2,\ldots$. 
Since $u_{jj}\equiv\sigma_{jj}(t\to\infty )$, 
we finally obtain that 
\begin{equation}
\sigma_{jj'}(t\to\infty)= {1\over N}\delta_{jj'}\ .
\label{a14}
\end{equation} 
This corresponds to the totally delocalized electron state. 
Since Eq.~(\ref{a14}) represents the unique solution of
Eqs.~(\ref{a11}), it implies that the asymptotic behavior 
of the electron density matrix is always given by Eq.~(\ref{a14})
for any initial conditions. Note that this result is true only
for $\Gamma\not =0$. Otherwise the solution of Eqs.~(\ref{a11})
is not unique.  

Eq.~(\ref{a14}) tells us that an arbitrarily weak interaction with 
the environment (detector) 
leads to delocalization in the Anderson model, 
even though this interaction affects only one of 
the sites. In other words, Anderson 
localization is unstable under infinitely small decoherence. 
One aspect of this instability is the importance of the order 
of limits $t\to\infty$ and $N\to\infty$. Taking $t\to\infty$ first,
as above, gives delocalization, while taking $N\to\infty$ first 
would preserve localization. In the non-interacting model, $\Gamma =0$,
the order of limits is immaterial and the electron is localized. 

Even though a local interaction with the environment destroys the 
localization, the latter should affect the   
time-dependence of the observed system. We expect  
the delocalization time to increase exponentially with $N$ 
and to be dependent on both the decoherence rate and the localization 
length. This matter deserves further investigation. 

We would like to stress that our result is not an effect of 
finite temperature, as is so called the hopping conductivity\cite{mott}. 
In the latter case, each site of the Anderson model interacts with the
thermal bath; in our case, only one site 
is coupled to the detector (environment). If we were to let    
all the sites interact equally with the detector ($I=I'$, Fig.~1), 
we would obtain no delocalization in our model, since $\Gamma=0$
in Eqs.~(\ref{a10}), (\ref{lind}) (see also\cite{gur1}).
Indeed, in this case Eq.~(\ref{lind}) is equaivalent to the Scr\"odinger
equation $i\partial_t|\Psi (t)\rangle 
=H_A|\Psi (t)\rangle$ leading to Anderson localization. 
Note that there is no measurement when $I=I'$.
The origin of delocalization in our case is therefore  
the break of coherence due to the measurement process.

Delocalization of the Anderson model due to measurement 
has been studied previously\cite{dittr,facchi,flores}. 
Yet the limit of a local and weak measurement 
has not been achieved. In the present work we include
the detector in the quantum mechanical description, avoiding the use
of the projection postulate in the course of measurement. This enables us 
to study delocalization due to local measurement and also in the limit
of weak coupling with the measurement device.

Another experimental setup for delocalization due to a 
local measurement is shown schematically in Fig.~3.
It can be realized in atomic systems, for instance, 
in experiments with Rydberg atoms\cite{uzi}. 
For $N=2$ this setup is similar to a V-level system used 
for investigation of the quantum Zeno effect\cite{zeno}.
The occupation of $E_1$ is  
\begin{figure} 
\psfig{figure=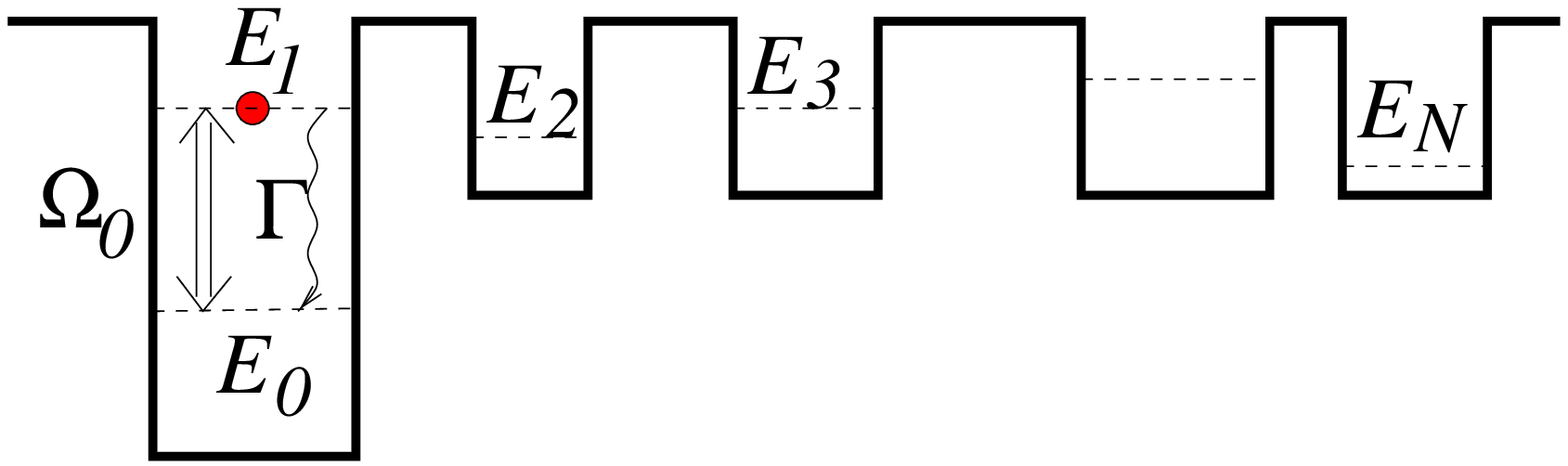,height=3cm,width=7cm,angle=0}
\noindent
{\bf Fig.~3:}
The 0-1 transition is driven by an intense laser field.
$\Omega_0$ and $\Gamma$ represent the Rabi frequency and 
the natural linewidth of the level $E_1$.   
\end{figure}
\noindent
 monitored via spontaneous 
photon emission, where the $0-1$ Rabi transition is generated by a laser 
field. 

Using the same derivation as 
in the previous case, Fig.~1, we obtain Bloch equations for 
the reduced electron density matrix $\sigma_{jj'}(t)$ where   
$j,j'={0,1,\ldots ,N}$. The off-diagonal density-matrix elements 
are described by the same 
Eq.~(\ref{a10b}). Equation (\ref{a10a}) for 
the diagonal density-matrix elements, however, is modified.
Now it reads
\begin{eqnarray}
&&\dot\sigma_{jj}=i\Omega_{j-1} (\sigma_{j,j-1}-\sigma_{j-1,j})
\nonumber\\
%&&~~~~~~~~~~~~~~~~~~~~~~~~~~~~~~~~~~~
&&~~~~~+i\Omega_j (\sigma_{j,j+1}-\sigma_{j+1,j})
-\Gamma(\delta_{j1}-\delta_{j0})\ \sigma_{11}\ .
\label{a15}
\end{eqnarray}
The last term in Eq.~(\ref{a15}) describes the rates due to spontaneous 
photon emission, Fig.~3. Here again the Bloch equations for the 
electron density matrix can be rewritten in Lindblad form,  
Eq.~(\ref{lind}), with $Q_{jj'}=\delta_{1j}\delta_{1j'}$ and 
$\tilde Q_{jj'}=\delta_{0j}\delta_{1j'}$.
(For $N=2$ Eq.~(\ref{lind}) coincides with the optical Bloch equations 
used for analysis of a V-level system\cite{fre}). Similar to the 
previous case, Fig.~1, Anderson localization is destroyed 
for any value of $\Gamma$, and the asymptotic electron distribution,
$\sigma_{jj}(t\to\infty )$, does not depend on the initial electron state. 
Here, however, the electron density matrix in the asymptotic state 
is {\em not} a pure mixture, $\sigma_{jj'}(t\to\infty )\not =0$, 
and the probabilities $\sigma_{jj}(t\to\infty )$ are not equally 
distributed between different wells (c.f. Eq.~(\ref{a14})).

The delocalization of the Anderson model should also 
affect its transport properties. Indeed, by connecting the first 
and the last dot in Fig.~1 to leads (reservoirs) 
one can expect current to flow through the dot array  
whenever any of the dots is monitored. Indeed, the 
stationary current through coupled dots is proportional to 
the occupation probability of 
the last dot, attached to the collector\cite{gur2}. 
The current should appear with a delay 
after a voltage bias to the leads is switched on. 
This time delay is precisely 
the relaxation time needed for the electron to be
delocalized. 

Anderson localization appears not only in  
quantum mechanics, but also in   
classical wave mechanics. Therefore the described delocalization 
due to local interaction with an environment should have 
a classical analogy. It can appear, for instance, in propagation 
of waves through coupled cavities with randomly distributed resonant
frequencies. A wave cannot ordinarily penetrate through such a system 
due to the Anderson localization. Random vibration 
of one of the cavities, however, should destroy the localization, so that  
waves begin to penetrate through the system after some time delay,
corresponding to the delocalization time. Such an experiment 
can also be done using the system of transparent plates 
with randomly varying thicknesses, described in\cite{bk}.

I am grateful to A. Buchleitner, B. Elattari, U. Smilansky
and B. Svetitsky for very useful discussions and important 
suggestions.

\end{multicols}
\end{document}